\documentclass[usenatbib,iop,numberedappendix]{aeb_emulateapj_2015}
\usepackage{amsmath}
\usepackage{amssymb}
\usepackage{xspace}
\usepackage[normalem]{ulem}
\usepackage{pifont}

\usepackage{natbib}

\usepackage[usenames,dvipsnames]{color}
\usepackage{hyperref}
\definecolor{darkblue}{rgb}{0.0,0.0,0.3}
\hypersetup{colorlinks,breaklinks,linkcolor=darkblue,urlcolor=darkblue,anchorcolor=darkblue,citecolor=darkblue}

\def\s{{\rm s}} 

\def\GHz{{\rm GHz}} 

\def\m{{\rm m}} 
\def\pc{{\rm pc}} 
\def\kpc{{\rm k}\pc} 
\def\Mpc{{\rm M}\pc} 

\def\erg{{\rm erg}} 

\def\mas{{\rm mas}} 
\def\muas{\mu{\rm as}} 

\def\m87{M87\xspace}                    

\begin{document}

\title{
Reconciling EHT and Gas Dynamics Measurements in M87: Is the Jet Misaligned at Parsec Scales?
}

\newcommand{\perimeter}{2}
\newcommand{\waterloo}{1}
\newcommand{\wcfa}{3}

\author{
  Britton~Jeter\altaffilmark{\waterloo,\perimeter,\wcfa},
  Avery~E.~Broderick\altaffilmark{\waterloo,\perimeter,\wcfa},
}
\altaffiltext{\waterloo}{Department of Physics and Astronomy, University of Waterloo, 200 University Avenue West, Waterloo, ON N2L 3G1, Canada}
\altaffiltext{\perimeter}{Perimeter Institute for Theoretical Physics, 31 Caroline Street North, Waterloo, ON N2L 2Y5, Canada}
\altaffiltext{\wcfa}{Waterloo Centre for Astrophysics, University of Waterloo, Waterloo, ON N2L 3G1, Canada}

\shorttitle{EHT Constraints on Jet Misalignment}
\shortauthors{Jeter et al.}

\begin{abstract}
The Event Horizon Telescope mass estimate for M87* is consistent with the stellar dynamics mass estimate and inconsistent with the gas-dynamics mass estimates by up to $2\sigma$.  We have previously explored a new gas-dynamics model that incorporated sub-Keplerian gas velocities and could, in principle, explain the discrepancy in the stellar and gas-dynamics mass estimate.  In this paper, we extend this gas-dynamical model to also include non-trivial disk heights, which may also resolve the mass discrepancy independent of sub-Keplerian velocity components.  By combining the existing velocity measurements and the Event Horizon Telescope mass estimate, we place constraints on the gas disk inclination and sub-Keplerian fraction.  These constraints require the parsec-scale ionized gas disk to be misaligned with the milliarcsecond radio jet by at least $11^{\circ}$, and more typically $27^{\circ}$.  Modifications to the gas-dynamics model either by introducing sub-Keplerian velocities or thick disks produce further misalignment with the radio jet.  If the jet is produced in a Blandford--Znajek-type process, the angular momentum of the black hole is decoupled with the angular momentum of the large-scale gas feeding M87*.
\end{abstract}


\section{Introduction}

Supermassive black holes (SMBHs) are thought to exist at the centers of most if not all galaxies in the universe, with masses ranging from $10^5 M_{\odot}$ to $10^{10} M_{\odot}$.  These massive objects are fundamentally important to their host galaxy, providing the engine to regulate star formation via highly energetic outflows.  The observed energy content of these outflows ranges from $10^{42} ~\erg ~\s^{-1}$ to $10^{48} ~\erg ~\s^{-1}$, and can only be reasonably produced by relativistic jets or winds thrown off SMBHs.  

Additional evidence of the existence of nuclear SMBHs comes from measurements of the velocity of stars or gas in the very centers of galaxies.  The velocity profiles measured in these observations require point source potentials with masses between $10^{6} M_{\odot}$ and $10^{10} M_{\odot}$, depending on the system, which can only be explained astrophysically with SMBHs.  The estimates for SMBH mass produce very tight correlations with galactic velocity dispersion, luminosity, and total mass, which is expected if SMBHs play such a critical role in regulating galaxy evolution \citep{Korm-Ho:13}. 

In 2017 April, a network of seven radio telescopes operating as a very long baseline interferometer (VLBI) observed the nuclear region of the giant elliptical galaxy Messier 87 as part of the Event Horizon Telescope \citep[EHT;][]{EHTM87-1:19, EHTM87-2:19, EHTM87-3:19}.  This observation produced the first image of the shadow of an SMBH, and measured the angular diameter of the ring-like emission to be $42 \pm 3 ~\muas$ \citep{EHTM87-4:19, EHTM87-6:19}.  Since the angular diameter of the black hole shadow is directly proportional to the black hole mass, one can estimate the mass of a black hole by measuring the angular diameter of the black hole shadow projected onto its surrounding material.  To make a precise mass estimate, there was an extensive effort to calibrate the angular gravitational radius measurement on simulated observations of hundreds of general relativistic magnetohydrodynamic (GRMHD) simulations \citep{EHTM87-5:19, EHTM87-6:19}.  Assuming a distance to M87 of $16.8~\Mpc$, the observed angular diameter corresponds to a black hole mass of $(6.5 \pm 0.7) \times 10^9 M_{\odot}$.  This mass estimate is consistent with the black hole mass estimate of $(6.6 \pm 0.4) \times 10^9 M_{\odot}$ made by modeling the kinematics of stars in the sphere of influence of the central black hole \citep{M87stars:11}, which modeled the gravitational potential of the entire M87 galaxy, including contributions from the central black hole mass and a cored dark matter distribution.  Most of the uncertainty in the EHT mass estimate comes from differences between the GRMHD simulations that were used as calibrators; the primary emission region can change significantly depending on the amount of magnetic flux (MAD/SANE) and the electron heating prescription \citep[${\rm R_{high}}$;][]{EHTM87-5:19}.   

There have been a few attempts to estimate M87's SMBH mass by measuring the dynamics of ionized gas in the nuclear region of M87, the most recent of which used the Hubble Space Telescope to measure the gas line-of-sight velocity to a precision on the order of $10~{\rm km ~s}^{-1}$ \citep{M87gas:13}.  To convert the velocity measurements into a black hole mass, the gas disk was assumed to be geometrically thin and inclined at $42^{+5} _{-7}$ degrees, and the gas disk was assumed to rotate with the Keplerian azimuthal velocity.  This model produces a black hole mass estimate of $3.5^{+0.9}_{-0.7} \times 10^{9} M_{\odot}$, which is discrepant with the EHT and stellar kinematics estimate by a factor of 2.   Even though the gas velocities were measured precisely, this dynamics model also systematically under-estimates the observed velocity dispersion by approximately $150 {\rm km ~s}^{-1}$.  Earlier observations of ionized gas in the central region of M87 also claim a black hole mass of $(3.2 \pm 0.9) \times 10^{9} M_{\odot}$ assuming a thin, Keplerian disk, but with less spatial resolution and significant uncertainty in the gas disk inclination, between $47^{\circ}$ and $65^{\circ}$ \citep{Macc-M87gas:97}.  

The black hole at the heart of M87 also has a prominent relativistic jet, observable at all wavelengths, and extending out $40~\kpc$ from the center.  This jet outputs a kinetic luminosity of approximately $10^{44} ~\erg ~\s^{-1}$ \citep{Bick-Begel:96, Sta-Aha-etal:06, Brom-Levin:09}, and contains multiple superluminal regions, reaching $6 c$ at a projected distance of $60 \pc$ \citep{Giro-Hada-etal:12}.  Such high apparent velocities place a strong constraint on the jet inclination of $\approx 18^{\circ}$. While some SMBH-launched relativistic jets in other galaxies show bends, kinks, or deflections, M87's jet is very straight out to a distance of about $10~\kpc$.  Close to the black hole, within $3 ~\mas$, a long campaign of radio observations at $43 \GHz$ constrains the jet viewing angle to be between $13^{\circ}$ and $27^{\circ}$ by leveraging the apparent velocities and brightness ratio of components in the forward and counter jet \citep{MLWH:16} (hereafter MLWH).  However, there is still some uncertainty in the radio-jet viewing angle across the field; earlier studies at $43 \GHz$ claimed a viewing angle between $30^{\circ}$ and $45^{\circ}$ \citep{Ly-Walk-Juno:07} using just the ratio of apparent velocities between the jet and counter jet.  Observations at $86 \GHz$ also estimate a jet inclination between $29^{\circ}$ and $45^{\circ}$, also based on apparent velocities \citep{Hada-Kino-Doi:16}.  Estimates of the jet inclination based on the brightness ratio between the jet and counter jet generally give lower inclination estimates \citep{Reid_etal:89} than apparent velocities, but the brightness ratio itself is difficult to estimate consistently across frequencies due to the low detectablility of the counter jet \citep{Hada-Kino-Doi:16}.  

In this paper, we consider the black hole mass estimate from the EHT 2017 observations  \citep{EHTM87-4:19,EHTM87-5:19,EHTM87-6:19} in concert with the historical gas-dynamics velocity measurements \citep{Macc-M87gas:97, M87gas:13} to explore the permitted regions of disk inclination and sub-Keplerian factor parameter space.  We also extend the gas-dynamics model presented in \citet{Jeter-Brod-McNam:19} to incorporate thick gas disks, and discuss the associated systematic on disk inclination.  We find that allowing for either sub-Keplerian gas velocities or non-trivial disk thicknesses moves the gas disk inclination to be more misaligned with the $43 \GHz$ radio jet compared to a thin, fully Keplerian disk, suggesting that the black hole angular momentum is decoupled from the angular momentum of the large-scale gas disk.  

\section{Implications for the mass and gas dynamics of \m87}

The measurement of a large black hole mass in \m87 has a number of implications for the relationship between the black hole and its immediate environment.  Large departures from non-axisymmetric motions would be difficult to reconcile with the nearly symmetric velocity profiles observed and theoretical expectations regarding the dynamics of gas flows, which are expected to efficiently symmetrize via viscous or magnetic interactions.  The observed line-of-sight velocities of atomic line emission produced within the gas flow at 10--100\,pc now places strong constraints on the dynamical state of gas on these scales and its relationship with the horizon and milliarcsecond-scale features.  The gas velocity's low apparent value may indicate significantly different inclinations than those inferred from dynamical modeling or substantial departures from circular Keplerian motions \citep[see, e.g.,][]{Korm-Ho:13,Jeter-Brod-McNam:19}.  Both of these are reasonable modifications to the pc-scale gas-flow models.

\begin{figure}
\begin{center}
\includegraphics[width=\columnwidth]{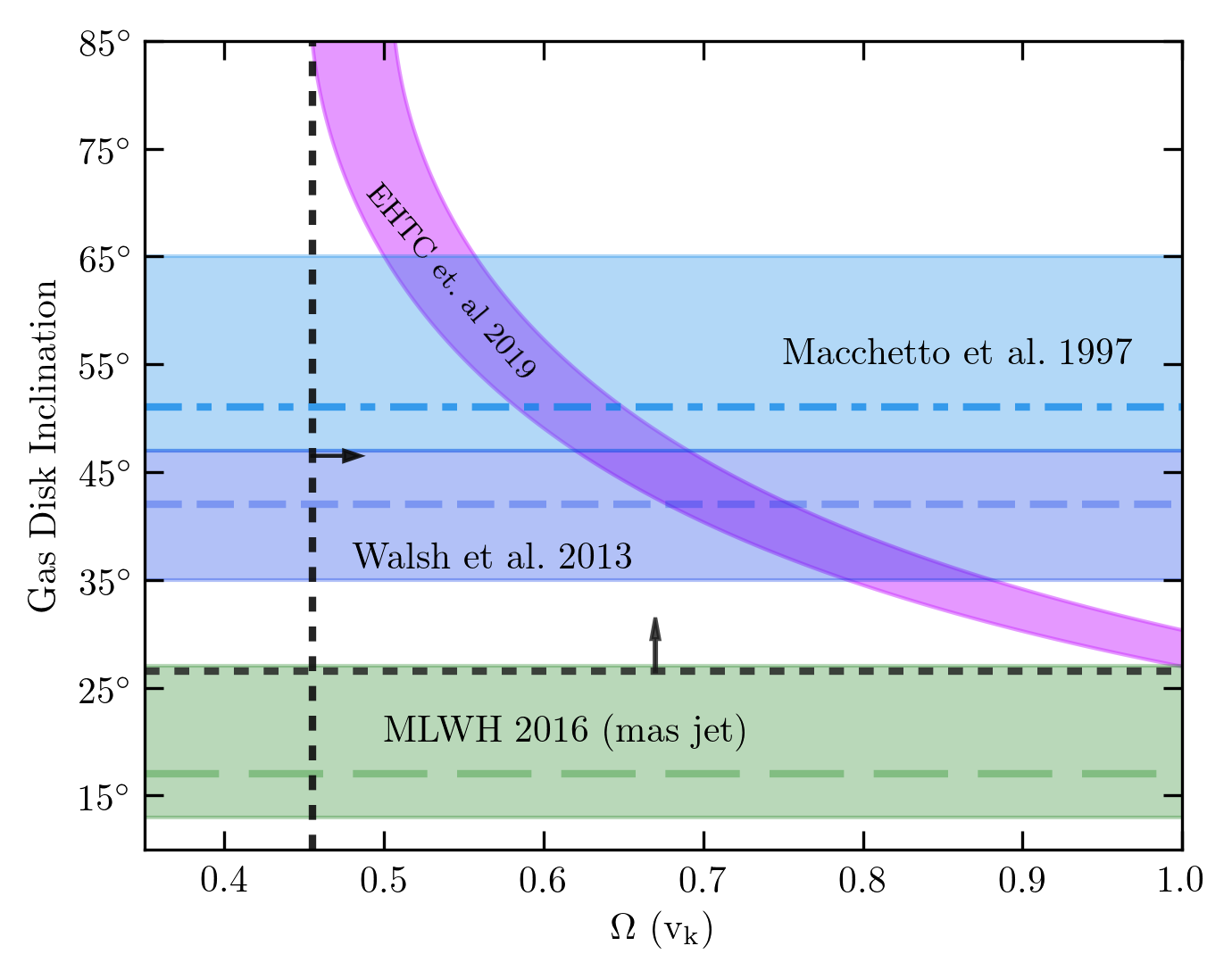}
\end{center}
\caption{Constraint imposed by $M/D$ measurement and observed gas velocities on the dynamics of the parsec-scale accretion flow in \m87.  The purple band indicates the range permitted by the EHT mass measurements when combined with the gas velocities in \citet{M87gas:13}.  The green horizontal long-dashed line shows the typical inclination of the mas-scale radio-jet, with green uncertainty bands; the blue (dashed) and light blue (dashed-dot) lines show the best-fit inclinations from the gas-dynamics measurements presented in \citet{M87gas:13} and \citet{Macc-M87gas:97}, respectively, and the colored bands show their uncertainty.  Requiring the gas disk to not exhibit super-Keplerian velocities imposes a lower limit on the disk inclination of $26^{\circ}$ (black dashed horizontal line), and requiring the gas disk to be inclined less than $90^{\circ}$ imposes a lower limit on the gas azimuthal velocity of $\approx 0.45 ~{\rm v_k}$ (black dashed vertical line).
} \label{fig:gas_flow}
\end{figure}

In \autoref{fig:gas_flow} we show the implied constraint on gas orbital velocity and inclination arising from a simple, axisymmetric gas disk model that permits sub-Keplerian velocities.  Note that because the Keplerian velocity at an angular displacement of $\theta$ is $\sqrt{GM/D\theta}$, this relationship depends on $M/D$ directly when the gas-flow velocity is scaled by the local Keplerian value.  The width of the band primarily arises from the uncertainty in the measurement of $M/D$ reported in \citet{EHTM87-6:19}, with a small contribution from the uncertainty in the gas velocity measurements from \citet{M87gas:13} added in quadrature.

There is no \textit{a priori} reason to believe that the large-scale gas flows must be aligned with features on horizon scales. There is evidence that the radio jet has been reasonably stable over many Myr \citep[see, e.g., Section 2 of][and references therein]{Brod-Nara-Korm:15}.  Nevertheless, at its present low accretion rate, even a highly misaligned accretion flow would require many orders of magnitude more time to substantially reorient the black hole spin.

Such a misalignment may be borne out in practice: even with the considerably larger black hole mass, it remains difficult to align the pc-scale gas flow with the inferred milliarcsecond-scale, and now horizon-scale, jet feature.  Inclinations less than $27^{\circ}$ require super-Keplerian flows.  This suggests that the gas-flow orientation is discrepant with the mas radio-jet by $\sim 10^{\circ}$.  It is notable that the PA of the gas flow of $45^{\circ}$ reported in \citet{M87gas:13} is consistent with the PA of the black hole spin implied by the ${\rm PA}_{\rm FJ}$ estimates in \citet{EHTM87-5:19,EHTM87-6:19}.  However, there is little direct evidence for these inclinations from direct observations of the gas disk itself.

Estimates of the gas disk inclination based on the geometry of the arcsecond-scale optical emission range from $47^{\circ}$ to $64^{\circ}$ \citep{Macc-M87gas:97} and $35^{\circ}$ to $47^{\circ}$ \citep{M87gas:13}.  These require significantly sub-Keplerian gas-flow velocities, ranging from 90\% down to 44\% of the Keplerian value.  This limit on the sub-Keplerian value is consistent with expectations from accretion models for low-luminosity active galactic nuclei, including non-radiative GRMHD simulations of gas flows onto black holes \citep[see, e.g.,][and references therein]{EHTM87-5:19}.  In these models, substantial pressure gradients within the flow provide additional radial support, and thus the orbital velocity and vertical disk structure are related via $v/v_K\propto \sqrt{1-h^2}$, where $h=H/R$ is the ratio of the accretion flow height and radial position, with typical values around $0.3$.  While these GRMHD models only claim to describe the innermost region ($R<100 ~R_g$) of the accretion flow, semi-analytic models of radiatively inefficient accretion flows (RIAFs) are self-consistent out to thousands of gravitational radii, and regularly produce similar disk thicknesses and sub-Keplerian values \citep{Quat-Nar:99}.

\subsection{Thick Disk Gas Models}

All of the existing ionized gas-dynamics mass estimates presume a thin, Keplerian disk.  Where in the previous paper, we describe relaxing the requirement of Keplerian motions, here we relax the condition that the disk is thin, motivated by results from simulations that show significant disk thicknesses out to large radii.

To model a thick disk, we follow the same procedure outlined in \citet{Jeter-Brod-McNam:19} to model ionized line emission with the addition of a vertical density profile.  However, instead of evaluating the line intensity at the disk equator, we evaluate along the line of sight to produce an integrated line intensity in the image plane.  Since the disk has a height, the line-of-sight velocity has an additional contribution in the vertical direction, 
\begin{equation}
  \vec{k} \cdot \vec{\beta}
  = - \frac{\cos i}{Rc} \alpha v_{k} Z +  ~\frac{\sin i}{Rc} \left( - \alpha v_{k} Y + ~\Omega v_{k} X \right),
  \label{eqn:k_dot_beta}
\end{equation}
where $i$ is the inclination of the disk; $X,Y$ and $Z$ are the Cartesian coordinates of the disk, oriented such that $X$ is parallel to a distant observer's x-axis, and $Z$ is aligned with the disk axis.  

We assume the disk has a vertical Gaussian emission profile $\rho$, with a scale height $H=hR$, where $h$
 is a fractional parameter that determines the disk thickness as a function of the radial distance $R$ from the center of the disk:
 \begin{equation}
  \rho=\frac{1}{\sqrt{2 \pi H^{2}}} \exp \left( -\frac{Z^{2}}{2H^{2}} \right).
  \label{eqn:vert_density}
\end{equation}
The ionized gas line shape is given by
\begin{equation}
  \Phi = \int g^{3}(z) \phi_{0}(z) \rho(z) dz
  \label{eqn:intensity_element}
\end{equation}
where, for a given height element $dz$, $g$ is the normal Doppler factor associated with the gas motion, $\phi_{0}$ is the natural line profile (which we assume to be Gaussian), and $\rho$ is the local density.  We adaptively integrate Equation \ref{eqn:intensity_element} along the line of sight to produce an integrated line shape.  We then multiply this line profile by the spatial emissivity profiles measured in \citet{M87gas:13} to construct a line intensity map in the image plane.  

\begin{figure}
\begin{center}
\includegraphics[width=\columnwidth]{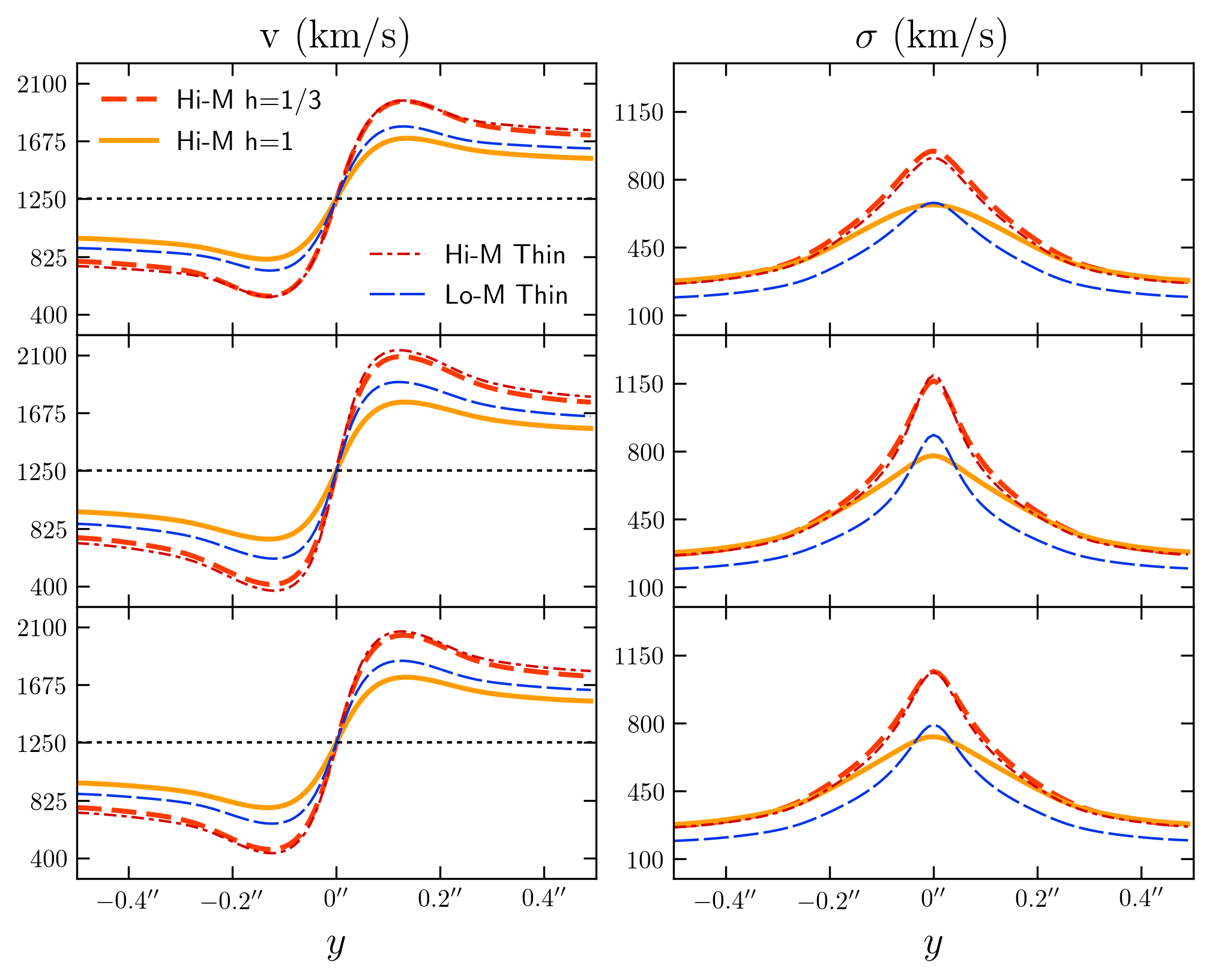}
\end{center}
\caption{Simulated velocity and dispersion profiles for three slits placed across the central 0$\overset{''}{.}$3 of M87.  All curves correspond to an optically thin Keplerian gas disk inclined at $42^{\circ}$, but with different thick disk scale heights.  The solid orange curve corresponds to a gas disk with scale height $h=1$, the dashed red curve to a gas disk with scale height $h=1/3$, the dashed-dot maroon curve to thin gas disk, and the long-dashed blue curve also to a thin gas disk.  Hi-M (High Mass) and Lo-M (Low Mass) correspond to a central black hole mass of $6.5 \times 10^9 M_{\odot}$ and $3.5 \times 10^9 M_{\odot}$, respectively.}
 \label{fig:disk_height}
\end{figure}

Similar to the procedure from \citet{Jeter-Brod-McNam:19}, we conduct simulated observations of the intensity map to produce velocity and dispersion profiles at different x-positions in the image plane, analogous to Hubble STIS slits.  The estimated line-of-sight velocity and dispersion curves are plotted in Figure \ref{fig:disk_height} for the three innermost slits, corresponding to slits with image plane x-positions of 0$\overset{''}{.}$1, 0", and -0$\overset{''}{.}$1, and spanning 0$\overset{''}{.}$5 to -0$\overset{''}{.}$5 in the image plane y-direction.  

For modest disk thicknesses of $h=1/3$ (thick, red, dashed line), similar to those seen in RIAF simulations at large radii, the line-of-sight velocity is suppressed by a few percent at all y-positions relative to a thin disk with the same black hole mass (thin, maroon, dashed-dotted line).  For larger thicknesses, like $h=1$ (thick, orange, solid line), the disk is quasi-spherical and the line-of-sight velocity is suppressed by nearly $50\%$ relative to the geometrically thin disk of the same black hole mass.  Large disk scale heights can even suppress the line-of-sight velocities well below that of a thin disk with a central black hole with half the mass.  Here, the larger mass (Hi-M) corresponds to the \citet{EHTM87-6:19} black hole mass estimate of $6.5 \times 10^{9} M_{\odot}$, and the lower black hole mass (Lo-M) corresponds to the \citet{M87gas:13} mass estimate of $3.5 \times 10^{9} M_{\odot}$.  

Because we assume the disk is optically thin and inclined toward the observer at $42^{\circ}$, the overall effect of increasing the disk height is to suppress the line-of-sight velocities.  In the forward half of the disk, closest to the observer, the line profiles from gas above the equatorial plane contributes more to the total line-of-sight velocity, which biases the gas velocities as if they were observed at a larger radius.  Similarly for the back half of the disk, away from the observer, gas below the equatorial plane contributes more to the total line-of-sight velocity, producing the same velocity bias toward larger radii.

\subsection{Velocity Bias in Geometrically Thick, Optically Thin Gas Disks}

The observed line-of-sight velocity, $v_{\rm obs}$, should be related to the intrinsic Keplerian orbital velocity $v_k$ by 
\begin{equation}
v_{obs} = \sin(i) F(i, h) \Omega v_k,
\label{eqn:F_corr}
\end{equation}
where $i$ is the inclination of the disk, $\Omega$ is the fractional intrinsic sub-Keplerian factor of the gas, and $F$ is the bias factor associated with the disk thickness.  This $F$ parameter should be dependent on both the disk inclination and the disk thickness.  

\begin{figure}
\begin{center}
\includegraphics[width=\columnwidth]{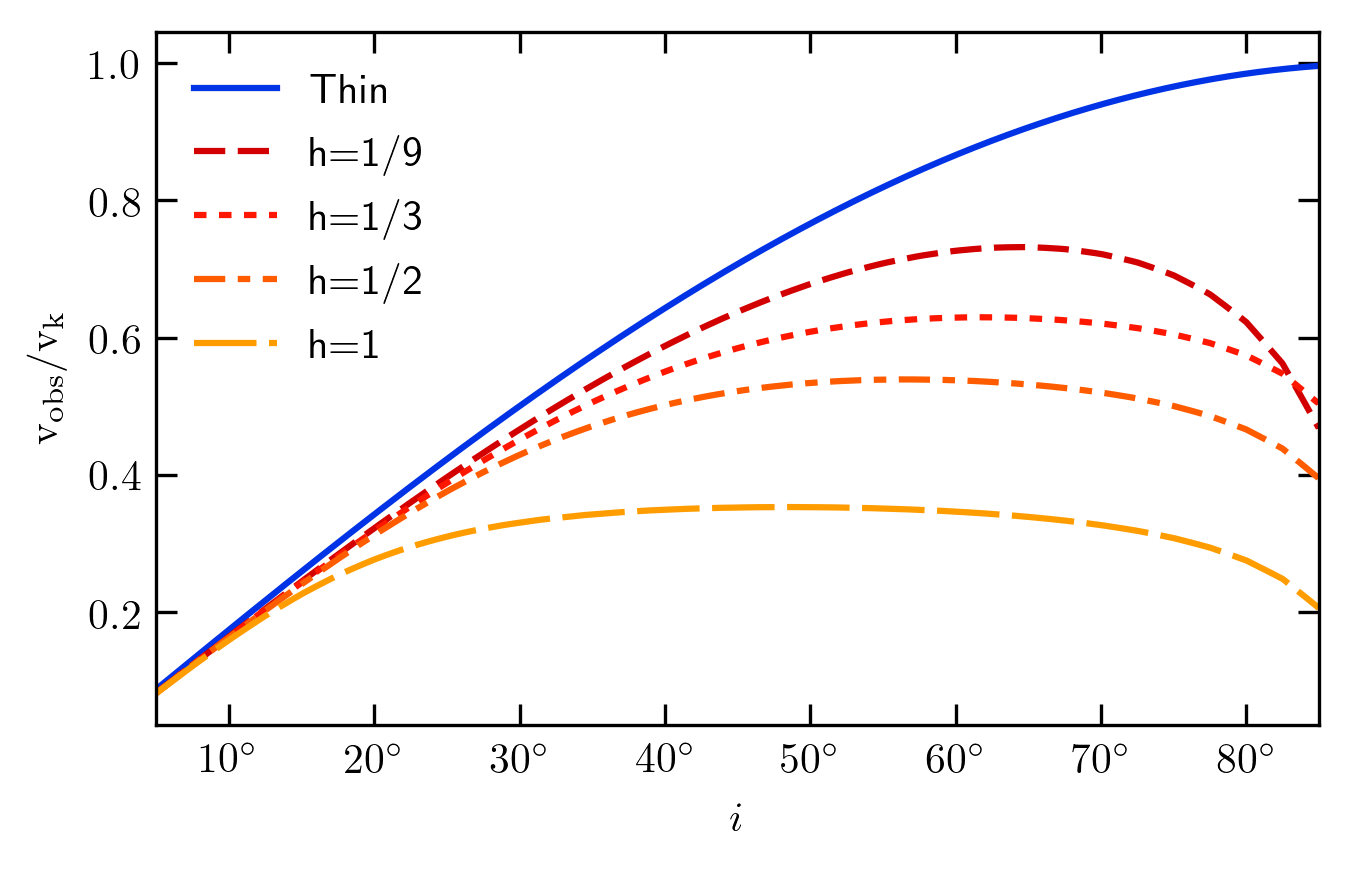}
\end{center}
\caption{Ratio of observed velocity to orbital Keplerian velocity as a function of disk inclination for ionized gas disks of different thicknesses, with no intrinsic radial velocity. 
} \label{fig:F_factor}
\end{figure}

In order to understand this bias factor $F$ in the observed velocity, we produced simulated observations of Keplerian ($\Omega=1$) gas disks with scale heights of $h=1/9, ~1/3, ~1/2, ~{\rm and} ~1$ with inclinations between $5^{\circ}$ and $85^{\circ}$, incremented every $5^{\circ}$.  We calculate the of the observed velocity to the orbital Keplerian velocity, and plot the result in Figure \ref{fig:F_factor}.  Thinner disks (h = 1/9, h = 1/3) provide less suppression of the observed velocity than thicker disks, but still produce reductions in the observed velocity on the order of $30-40 \%$.  Very thick disks (h = 1) provide significant suppression, upward of $70\%$.  As the disk becomes more face-on (inclination approaches zero), the suppression factor from the intrinsic inclination dominates the observed velocity.  

\begin{figure}
\begin{center}
\includegraphics[width=\columnwidth]{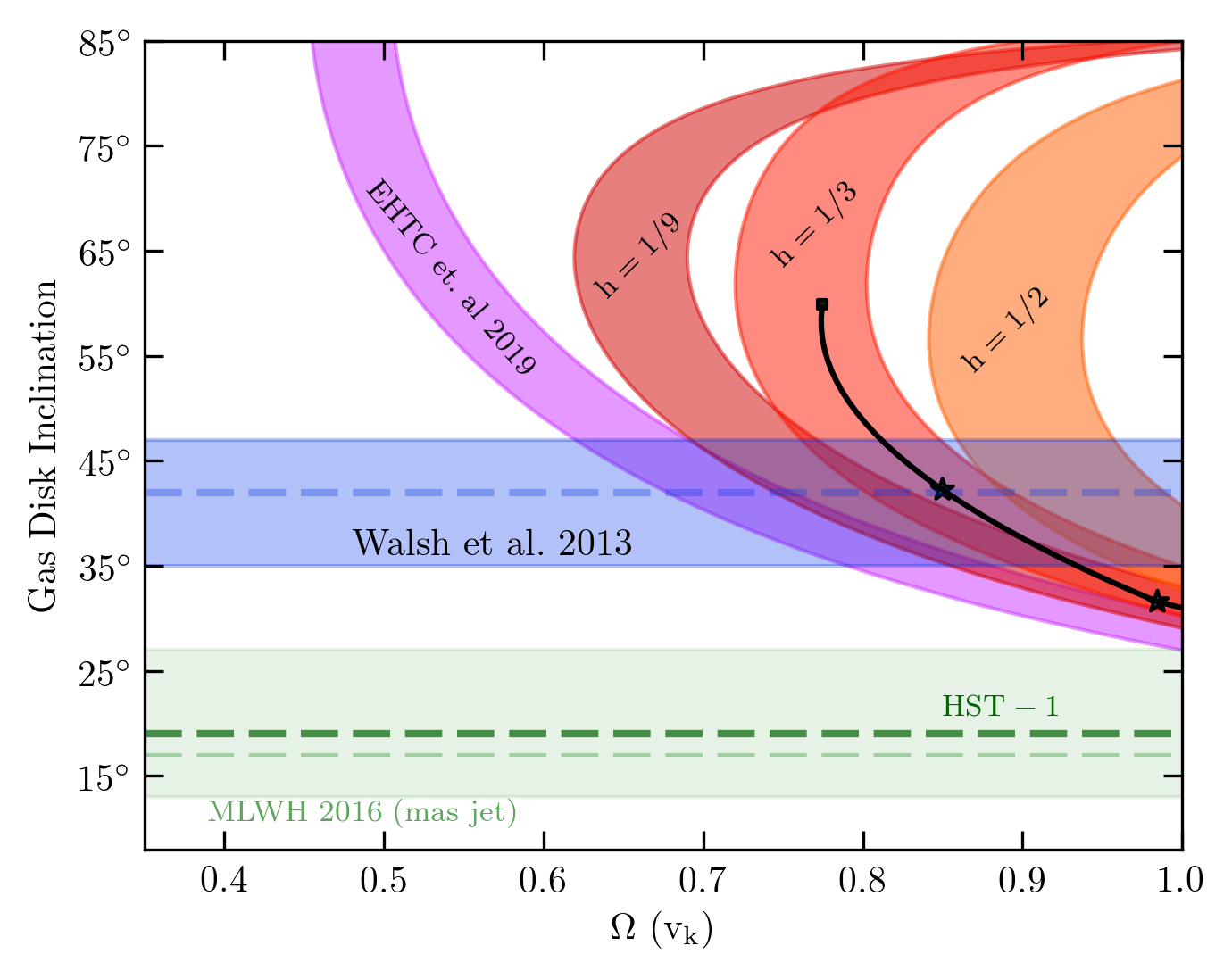}
\end{center}
\caption{Similar to Fig. \ref{fig:gas_flow}, now with dark-red (h = 1/9), red (h = 1/3) and orange (h = 1/2) bands showing the effects of non-trivial disk height on the implied inclination and sub-Keplerian factor of the ionized gas disk.  A disk with h = 1 requires super-Keplerian intrinsic velocities to be consistent with the EHT mass estimate.  The thick green dashed line represents the $19^{\circ}$ inclination limit implied by the $6c$ superluminal motion at HST-1.  The black curve represents the theoretical relationship between $\Omega$ and disk scale height inferred from RIAF solutions.  
} \label{fig:gas_flow_h}
\end{figure}

We can then take these observed velocity vs. inclination profiles and plot thick disks as different bands in the inclination/sub-Keplerian space, as shown in Figure \ref{fig:gas_flow_h}.  Here we remove the inclination band from \citet{Macc-M87gas:97} for clarity, and show the effective inclination for a disk with $h=1/9$ as the dark-red shaded region, $h=1/3$ as the red shaded region, and $h=1/2$ as the orange shaded region.  Since thick gas disks suppress the observed velocity in a similar way to inclination, a thick gas disk can have a larger inclination (more edge-on) than a thin gas disk and still maintain relatively Keplerian gas velocities.  Very thick gas disks (h = 1) require super-Keplerian values of at least $\Omega > 1.3 ~v_k$ for the observed line-of-sight velocities from \citet{M87gas:13} to be consistent with the EHT mass estimate.  Note that thick disks imply the gas disk inclination is generally more misaligned with the radio jet than thin disks.  The inclination implied by the $6c$ super-luminal motion at HST-1 \citep{BirettaHST1:99} is also marked as a thick green dashed line, and serves as an upper limit on the jet inclination at approximately $200 ~\pc$ from the black hole, assuming the intrinsic jet velocity is very relativistic ($\Gamma > 15$). 

 We do not extend the horizontal axis of this figure to super-Keplerian values because it is generally challenging to construct astrophysical scenarios where substantially thick disks produce super-Keplerian velocities.  The usual mechanisms to produce thick disks, like additional pressure, generally reduce the azimuthal velocities below the Keplerian values.  For example, in typical RIAF self-similar solutions, the disk thickness and sub-Keplerian factor are strongly related \citep{Quat-Nar:99}.  From the flow equations, one can find that the sub-Keplerian fraction $\Omega$ is related to the disk height by
\begin{equation}
\Omega = \frac{v}{v_k} = \left( 1 - \frac{5}{2} h^2 \right)^{1/2}, 
\end{equation}
where $v_k$ is the Keplerian azimuthal velocity and the factor of $5/2$ comes from the radial density profile.  This relationship is plotted as a black curve in Figure \ref{fig:gas_flow_h}, along with starred points corresponding to disk heights of $h=1/9$ and $h=1/3$.  The curve terminates with a black square at a disk height of $0.4$, which is where the RIAF relationship is no longer consistent with the observed velocity and mass measurements.   Note that a RIAF model with a disk height of one-third is consistent with the best-fit inclination from \citet{M87gas:13}.  

\section{Discussion}

\subsection{ Jet-disk Misalignment}

For thin disks, the parsec-scale disk inclination must be misaligned with the inclination of the high-frequency radio jet by at least $1\sigma$.  When allowing for either sub-Keplerian velocities, or geometrically thick disks, the misalignment between the gas disk inclination and the radio jet becomes more significant.  

Without invoking thick disks, the combined EHT mass estimate and \pc-scale gas velocity profile suggest that the large-scale gas is misaligned with the large scale jet by at least $15^{\circ}$ when using the typical values for jet inclination and black hole mass.  When also considering the azimuthal angle with respect to the line of sight, or position angle, the absolute misalignment may be larger.  The position angle of the gas disk major axis is approximately $45^{\circ}$ east of north \citep{M87gas:13}, which means the axis of rotation for the gas disk has a position angle of $315^{\circ}$ east of north.  The jet position angle is measured to be $288^{\circ}$ in \citet{Walk-Hardee-Davies:18}.  For the best case values of the jet and disk inclination, $25^{\circ}$ and $27^{\circ}$ respectively, the inner product of the jet axis and disk axis produces an absolute misalignment of at least $11^{\circ}$.  Again, increasing the disk thickness allows the disk to have higher intrinsic inclinations, and thus increases the absolute misalignment even further.  For example, taking the jet inclination measured at HST-1 \citep[$18^{\circ}$;][]{BirettaHST1:99}) and the disk inclination measured by \citet{M87gas:13} yields an absolute misalignment of $27^{\circ}$.  

Such a misalignment implies that the angular momentum of the large-scale gas is different from the angular momentum of the large-scale jet.  If the jet angular momentum is coupled to the black hole, the Bardeen--Peterson effect should align the inner accretion disk with the black hole angular momentum, but is only effective at much smaller radii, very close to the black hole, and highly dependent on the surface density and viscosity prescriptions for the accretion disk \citep{Bard-Petter:75, Natar-Pringle:98}.  The Lense--Thirring precession timescales for low-luminosity systems like M87 can be as high as a Gyr, and again depend on the density and viscosity structure of the disk \citep{Natar-Armit:99}.  This means a jet coupled to the black hole angular momentum may not necessarily be aligned with the angular momentum of the outer gas disk

Thick accretion disks generally exhibit enough dynamical support to prevent alignment away from the innermost stable circular orbit (ISCO) region, and end up producing solid-body-like rotations around the central black hole \citep{Frag-Annin:05,Dexter-Frag:11}.  In thick disk simulations where the disk is initially tilted with respect to the black hole the jet also quickly decouples from the angular momentum of the black hole, and becomes aligned with the inner sub-$100 ~ R_g$ accretion disk \citep{Chatterjee_Jets:20}.  In this scenario, the jet angular momentum is coupled to the inner-disk angular momentum.  The problem now is to investigate why and where the inner disk becomes misaligned with the outer disk. Such misalignments could arise from an asymmetrical mass distribution of the large-scale gas, or as a consequence from an asymmetric galactic potential.  Such an effect would be small in M87, since it has a relatively spherically symmetric radial stellar profile in the inner few dozen parsecs \citep{M87stars:11}, as well as a low-velocity anisotropy \citep{Zhu-Long-etal:14} at larger, kpc distances.  

\subsection{Applicability to Other Gas-Dynamics Techniques}

Much of the analysis in this work and in \citet{Jeter-Brod-McNam:19} concerns possible systematics when converting ionized gas velocity and dispersion measurements to black hole mass estimates.  Many recent studies have made black hole mass estimates using molecular gas velocity measurements using instruments like ALMA to great success.  Since ALMA observations are able to measure gas velocities at every resolution element, one can fully map out the velocity structure of circumnuclear molecular gas disks.  This is in contrast with the slit spectra method used in ionized gas measurements, where one axis (typically the x-axis) has a much lower resolution than the other axis in the image plane, compromising our ability to map the full spatial velocity structure.  

In well resolved ALMA systems, the clues that would point toward a more complicated velocity model generally do not exist.  These systems generally have very weak to nonexistent AGN activity, their velocity dispersions are typically much lower than in ionized gas systems, and in many cases, there exists a clear well-structured gas torus at other observing frequencies.  All these various features suggest that the molecular gas disks in these systems more closely trace the underlying black hole potential without needing significant virial components in their velocity model, and the molecular gas velocities are very well described by thin, Keplerian models.  

A common addition to both ionized and molecular gas-dynamics velocity models is a tilted or warped gas disk, generally invoked to explain kinematic residuals or "twists" in the innermost regions of the large-scale gas disk.  These warped disks are typically motivated by an asymmetric or perturbed galactic potential, but efforts to find either the physical source of such asymmetries or quantitatively estimate the amount of warp in the inner disk have not been successful.  Since warped disks appear in practice as a change in disk inclination with radius, one could imagine altering the disk thickness with radius to achieve a similar effect.

Megamasing disks around AGN are also famously well described by thin, Keplerian molecular gas disks, but may also contain warped disks \citep{MegaMas-10:18, MegaMas-11:20}.  These warps are similarly motivated by perturbed galactic potentials, but they could also arise from a dynamic, and small, radial velocity component that changes with radius.  If the disk was tilted by only a few degrees, with a small but non-negligible thickness, a small radial velocity component can produce what looks like a warped masing disk in the image plane, since the addition of radial velocities changes the preferred masing direction.  

\section{Conclusions}

The EHT mass estimate of M87* is inconsistent with mass estimates derived from ionized gas-dynamics models using thin, Keplerian gas disks.  By combining the EHT mass estimate and the velocity measurements from \citet{M87gas:13}, one can produce a combination of gas disk inclinations and sub-Keplerian factors that resolve the tension between the horizon-scale and ionized gas mass estimates.  Of note, there are no combinations of gas disk inclination and sub-Keplerian factor that, to within $1\sigma$, align the axis of the ionized gas disk with the axis of the large-scale radio jet.  The minimum gas disk inclination is $27^{\circ}$ with a fully Keplerian gas disk, and requires super-Keplerian velocities to breach the inclination limits imposed by HST-1 \citep{BirettaHST1:99}.  Using typical values for the jet ($18^{\circ}$) and disk ($42^{\circ}$) inclinations yields an absolute misalignment of $27^{\circ}$.

Invocations of sub-Keplerian factors or different inclinations are not the only ways to explain the apparent suppression of the line-of-sight gas velocities in M87.  The incorporation of non-trivial disk height can also produce a significant systematic when estimating black hole masses.  As long as the gas disk is optically thin and inclined with respect to the observer, an increased disk height serves to bias line-of-sight velocities to larger radii, suppressing the effective line-of-sight velocity.  This suppression produces a higher intrinsic disk inclination, and with disk heights of $h ~\approx ~1/3$ one can just resolve the discrepancy between the EHT and gas-dynamics mass estimates.  Any non-trivial amount of disk thickness tends to further misalign the large-scale gas disk with the large-scale jet.  

Achieving precise and accurate black hole mass estimates are critical to understanding the relationship between AGN feedback and galactic evolution.  ALMA black hole mass estimates have improved dramatically over earlier ionized gas mass estimates, but well-structured molecular disks appear in only $10\%$ of galaxies.  In order to improve our catalog of black hole masses, it is natural that a gas-dynamics model with the ability to investigate sub-Keplerian and non-trivial disk heights should be considered.  

\acknowledgments
We thank Kazuhiro Hada, Ramesh Narayan, and the anonymous referee for their helpful comments that improved this paper.  This work was supported in part by the Perimeter Institute for Theoretical Physics. Research at Perimeter Institute is supported by the Government of Canada through the Department of Innovation, Science and Economic Development Canada, and by the Province of Ontario through the Ministry of Research, Innovation and Science.
A.E.B. thanks the Delaney Family for their generous financial support via the Delaney Family John A. Wheeler Chair at Perimeter Institute.
B.J. and A.E.B. receive additional financial support from the Natural Sciences and Engineering Research Council of Canada through a Discovery Grant.

\bibliographystyle{apj}
\bibliography{m87}
\clearpage
\end{document}